\renewcommand{\a}{\alpha} \renewcommand{\b}{\beta}
\renewcommand{\d}{\delta}
\newcommand{\e}{\varepsilon} \newcommand{\z}{\zeta}
\newcommand{\h}{\eta} 
\renewcommand{\l}{\lambda}
\newcommand{\m}{\mu} \newcommand{\n}{\nu}
\newcommand{\x}{\xi}  \newcommand{\r}{\rho}
\newcommand{\s}{\sigma} 
\renewcommand{\o}{\omega}
 \newcommand{\ph}{\varphi}
 
\renewcommand{\L}{\Lambda}

\newcommand{\CA}{{\cal A}}

 \newcommand{\CH}{{\cal H}}
 
 \newcommand{\CL}{{\cal L}}
 
\newcommand{\CO}{{\cal O}}

\newcommand{\CU}{{\cal U}}

\newcommand{\6}{\partial}

\newcommand{\then}{\Rightarrow}
\newcommand{\lra}{\leftrightarrow}
\newcommand{\pr}{\prime}
\newcommand{\diag}{\mbox{\rm diag}}

\newcommand{\re}{\mbox{{\sf I \hspace{-0.8em} R}}}
\newcommand{\la}{\langle} \newcommand{\ra}{\rangle}

\documentstyle[12pt]{article}

\begin{document}
\begin{titlepage}
\renewcommand{\baselinestretch}{1.3}
\begin{flushright}
TUW - 92 - 04 \\
UWThPh - 1992 - 17
\end{flushright}
\vfill

\begin{center}
{\Huge Novel Symmetry of Non-Einsteinian
Gravity in Two Dimensions}
\vfill

{\Large Harald Grosse $^{1}$, Wolfgang Kummer $^{2}$,
Peter Pre\v{s}najder~$^{3}$ \\
and
Dominik J. Schwarz $^{2}$\footnote{e-mail:
dschwarz@email.tuwien.ac.at}}
\\ \medskip

$^{1)}$ Institut f\"ur theoretische Physik \\
Universit\"at Wien \\
Boltzmanngasse 5, A-1090 Wien \\
Austria \\ \smallskip

$^{2)}$ Institut f\"ur theoretische Physik \\
Technische Universit\"at Wien \\
Wiedner Hauptstra\ss e 8-10, A-1040 Wien \\
Austria\\ \smallskip

$^{3)}$ Matematicko-fyzik\'alna fakulta \\
Univerzita Komensk\'eho \\
Mlynsk\'a dolina F 2, CS-842 15 Bratislava \\
Czechoslovakia
\end{center}
\vfill

\begin{abstract}
The integrability of $R^2$-gravity with torsion in two
dimensions is traced to an ultralocal dynamical symmetry of constraints
and momenta in Hamiltonian phase space. It may be interpreted as
a quadratically deformed $iso(2,1)$-algebra with the deformation
consisting of the Casimir operators of the undeformed algebra.
The locally conserved quantity encountered in the explicit
solution is identified as an element of the centre of this
algebra.  Specific contractions of the algebra are related to
specific limits of the explicit solutions of this model.
\end{abstract}
\vfill

May 1992 \hfill \\
\pagebreak
\end{titlepage}

\section{Introduction}

The beautiful properties of string theory, interpreted as
Einstein-gravity in two dimensions with ``world-coordinates'' as
additional fields, are closely related to the Weyl-invariance
which arises as an additional ``accidental'' symmetry in two
space-time dimensions. Recently also a non-Einsteinian model of
gravity --- without Weyl-symmetry --- in two dimensions has
attracted interest \cite{kat1,kat2,ks1,ks2}, because its
integrability also led to the conjecture of possessing some
further symmetry \cite{ks1}.  It belongs to the variety of
gravitational theories with higher powers of curvature and with
torsion in the action \cite{hehl,schmidt}.  Restricting the
field equations to contain at most second order derivatives of
the field variables (zweibein $e_\m^{\ a}$ and spin-connection
$\o_{\m\ b}^{\ a}$), the action invariant under diffeomorphisms
and local Lorentz transformations is essentially unique
\begin{equation}
\label{inv}
L = -\frac{1}{4M^2} \int d^2\!x e ( R^{\quad ab}_{\m\n}
                  R^{\m\n}_{\quad ab} + M^2\b T^{\quad a}_{\m\n}
                  T^{\m\n}_{\quad a} + 4M^2\l )
\end{equation}
with
\begin{equation}
\label{rt}
\begin{array}{rcl}
e &=& \det(e_\m^{\ a}) \\
R^{\quad a}_{\m\n\ b} &=& \left( \6_\m \o_\n -\6_\n \o_\m \right) \e^a_{\ b}
                          =: F_{\m\n}\e^a_{\ b} \\
T^{\quad a}_{\m\n}    &=& \6_\m e^{\ a}_\n + \o_\m \e^a_{\ b} e^{\ b}_\n
                            - (\m \lra \n )
\end{array}
\end{equation}
where
\begin{equation}
\begin{array}{rcl}
\o^{\ a}_{\m\ b} &=:& \o_\m\e^a_{\ b} \\
(\e^{ab}) &=& \left(\matrix{ 0 & -1 \cr 1 & 0 \cr}\right)\; .
\end{array}
\end{equation}
Greek indices are lowered and raised by the metric
$g_{\m\n}=e_\m^{\ a} e_\n^{\ b}\h_{ab}$ and its inverse
$g^{\m\n}$, where $(\h_{ab})=\diag(+,-)$ is the Lorentz metric
operating with the Latin indices.  The (topological)
Einstein-Hilbert term $eR$, a total divergence, is omitted in
(\ref{inv}), $M$ and $\b$ are free parameters, $\l$ represents a
cosmological constant. $M,\b$ and $\l$ are chosen to
have mass dimension one, zero and two, in order to yield a
dimensionless action $L$. For simplicity we shall set $M^2 \equiv 8$
in the following.

The classical solutions of (\ref{inv}) were studied extensively
in the conformal gauge $e_\m^{\ a} = e^\ph \d^a_\m$ with $\ph$
and $\o_\m$ as dynamical variables. Complete integrability was
found \cite{kat1}. The solution consists of two branches, one
with constant curvature and vanishing torsion, the other one
with nontrivial curvature and torsion. The former is closely
related to Liouville theory. Recently all universal coverings of
geodesically complete solutions have been classified
\cite{kat2}.

The Euclidian version of the model (\ref{inv}) has also been
identified with the appropriate geometric formulation of
continuous distributions of dislocations and disclinations in
a two-dimensional membrane. The three-dimensional version of
this equation has been used for the treatment of elastic
media in Euclidian $d=3$ \cite{kat3}.

On  the other hand, (\ref{inv}) closely resembles a (noncompact)
gauge theory. This suggests also light-cone (LC) components and
the use of a LC gauge \cite{ks1,ks2}.  Working in LC coordinates
\[
x^\pm = x^0 \pm x^1
\]
simplifies the calculations considerably.
Introducing
\[
\begin{array}{rcl}
\o_{\pm} &=& \frac{1}{2}(\o_0 \pm \o_1) \\
e_{\pm}^{\ a} &=& \frac{1}{2}(e^{\ a}_0 \pm e^{\ a}_1)
\end{array}
\]
and grouping the field variables as ($i = 1,2,3$)
\begin{equation}
\label{q}
\begin{array}{rcl}
q_i &=& \left( \o_-, e_-^{\ -}, e_-^{\ +} \right) \\
\bar q_i &=& \left( \o_+, e_+^{\ -}, e_+^{\ +} \right),
\end{array}
\end{equation}
(\ref{inv}) can be rewritten as
\begin{equation}
\label{lcinv}
\CL = -\frac{1}{2e}(F_{+-})^2+\frac{2\b}{e} T_{+-}^{\quad +}
              T_{+-}^{\quad -} - e\l\; ,
\end{equation}
where
\begin{equation}
\label{16}
\begin{array}{rcl}
e &=& q_2\bar q_3 - q_3 \bar q_2 \\
F_{+-} &=& \dot q_1 - \bar q_1^\pr \\
T_{+-}^{\quad +} &=& \dot q_3 - \bar q_3^\pr + \bar q_1 q_3 -
q_1 \bar q_3 \\
T_{+-}^{\quad -} &=& \dot q_2 - \bar q_2^\pr - \bar q_1 q_2 +
q_1  \bar q_2\; .
\end{array}
\end{equation}
Dot and prime indicate the derivative with respect to $x^+$ and
$x^-$.  The homogeneous LC gauge is characterized by $\bar q_1 =
\bar q_2 = \bar q_3 - 1 = 0$ (cf.~(\ref{q})). It should be
emphasized that it can always be reached by appropriate
diffeomorphisms and local Lorentz transformations; the proof is
straightforward. In this gauge, as shown in \cite{ks1}, the
complete integral of the equations of motion can be even
expressed in terms of elementary functions of $x^+$. It was
found that the solutions depend on three arbitrary functions of
$x^-$ and one constant, the latter being related to the quantity
\begin{equation}
\label{Q}
Q = \exp (-{p_1\over 2\b}) \left( {E\over 2\b^2} + {p_1\over \b}
+ 2 \right)
\end{equation}
with
\begin{equation}
\label{E}
E = {p_1^2\over 2} - {p_2p_3\over 2\b} - \l\;.
\end{equation}
$p_i$ are the ``conjugate momenta'' to the ``coordinates'' $q_i$
and will be defined in Sec. 2.1. $p_1$ is proportional to the
curvature scalar, $p_2p_3$ to the torsion scalar, the second
term in (\ref{inv}).  Thus (\ref{Q}) behaves as a scalar under
diffeomorphisms and local Lorentz transformations and therefore
holds in any gauge.  $Q$ is constant for any solution, i.e.~it
is independent of both variables
\begin{equation}
\label{19}
\dot Q = Q^\pr = 0\; .
\end{equation}
The action (\ref{inv}) is nonpolynomial and thus its
renormalizibility is by no means evident, the more so, because
the usual specific infrared problems of $d=2$ are relevant
also here. However, as shown by two of the present authors
\cite{ks2}, at least in a flat background the path-integral
quantization can be carried through, and the theory turns out to
be renormalizable by fixing one (UV-divergent) constant in front
of an infinite number of counterterms.

The purpose of our present work is to investigate the symmetry
properties which are responsibe for the integrability  of
(\ref{inv}). Constants of motion like the ``local'' one
(\ref{Q}) must follow from this symmetry.

This symmetry can be found starting from the Hamiltonian
formulation. As is well-known, in the latter the algebra of all
first class constraints provides an opportunity to identify such
a symmetry \cite{hen1}. We show in Sec.~2.1 that the Hamiltonian
can be expressed completely in terms of the secondary
constraints. This could have been expected for a theory of
gravitation.  Actually the constraints resemble the ones in the
Ashtekar formulation of general relativity, being polynomials in
$p$ and $q$ \cite{ash}.

In Sec.~2.2 and 2.3 we solve the system of Hamiltonian equations
of motion, introducing in a natural manner conserved quantities
like (\ref{Q}).  Calling the evolution parameter $x^+$ ``time'',
the equations for the time-evolution of the momenta (Sec.~2.2)
and for the coordinates (Sec.~2.3) can be solved. The general
solution depends on three arbitrary functions of ``space''
$x^-$.  Exploiting the residual gauge-symmetry after fixing the
LC gauge it is possible to introduce a ``normalized'' solution,
were the arbitrary functions are replaced by constants
(Sec.~2.4). This shows that the $x^-$-dependence is essentially
irrelevant for the dynamics.

After this preparation we relate the different branches of
solutions to the dynamical symmetry (Sec.~3) which is the main
issue of our present work. This symmetry appears because the
algebra of secondary constraints in our present case closes
together with the momenta $p$. The algebra is nonlinear and can
be considered as a deformed $iso(2,1)$, with the deformation
consisting of the Casimir operators of the undeformed
$iso(2,1)$. The quantity (\ref{Q}) appears as one of the two
elements of the centre of this algebra.

\section{Hamiltonian Formulation}

\subsection{Hamiltonian}

Choosing $x^+$ to be the evolution parameter in the Hamiltonian,
the canonical momenta corresponding to (\ref{lcinv}) in LC
coordinates become ($p_i = {\d L\over \d\dot q_i}$)
\begin{eqnarray}
p_1 &=& - {1\over e} \left( \dot q_1 - \bar q_1^\pr \right) \nonumber \\
\label{p}
p_2 &=& {2\b\over e} \left( \dot q_3 - \bar q_3^\pr + \bar q_1 q_3
- q_1\bar q_3 \right) \\
p_3 &=& {2\b\over e} \left( \dot q_2 - \bar q_2^\pr - \bar q_1 q_2
+ q_1 \bar q_2 \right) \nonumber
\end{eqnarray}
with primary constraints
\begin{equation}
\label{pc}
\bar p_i = {\d L\over \d\dot{\bar q}_i} \sim 0\; .
\end{equation}
We define the Poisson brackets ($x^+=y^+$):
\begin{equation}
\label{pb}
\begin{array}{rcl}
[ q_i(x),p_j(y) ] &=&  \d(x^- - y^-)\d_{ij} \\
{[} {\bar q}_i(x),{\bar p}_j(y) ] &=& \d(x^- - y^-)\d_{ij} \\
{[} q_i(x),q_j(y) ] &=& [ p_i(x),p_j(y) ] = \dots =0
\end{array}
\end{equation}
With the canonical Hamiltonian $H_C = \int dx^- ( \dot q_i p_i - \CL ) =
\int dx^- \CH_C$ they entail secondary constraints
\begin{equation}
G_i := \dot{\bar p}_i = - {\6 \CH_C \over \6\bar q_i} \sim 0\; ,
\end{equation}
where
\begin{equation}
\label{sc}
\begin{array}{rcl}
G_1 &=& p_1^\pr + q_3p_3 - q_2p_2 \\
G_2 &=& p_2^\pr + q_1p_2 - q_3E \\
G_3 &=& p_3^\pr - q_1p_3 + q_2E
\end{array}
\end{equation}
with $E$ already defined in (\ref{E}). The Hamiltonian turns out
to be linear in $G_i$
\begin{equation}
\label{H}
H_C (x^+) =\int dx^- \left( P^\pr - \bar q_i G_i \right)\; ,
\end{equation}
where
\begin{equation}
\label{P}
P := \bar q_ip_i\; .
\end{equation}
In the course of our study of the algebra of $G_i$ below
(Sec.~3), we shall find that the $G_i$ are in involution. This
means the absence of ternary constraints, all primary and
secondary constraints are first class.

Since there is no $\bar p_i$-dependence in the Hamiltonian, the
canonical Heisenberg equations of motion for $\bar q_i$ imply
$\dot{\bar q}_i = [\bar q_i,H_C] = 0$.
If one uses the formalism of extended Hamiltonian \cite{hen}
$\CH_E = \CH_C + \m_i\bar p_i + \l_iG_i$, where all first class
constraints are added through Lagrange multipliers $\m_i$ and
$\l_i$, the analogous equation reads $\dot{\bar q}_i = \m_i$.
$\m_i$ are unphysical fields and may be ``gauge fixed'', i.e.~we
will work at $\l_i=\m_i=0$ throughout this paper (the $\l_i$ are
``absorbed'' by the $\bar q_i$ in $H_C$). This leads exactly to
the class of general LC gauges ($\bar q_i = \bar q_i (x^-)$) of
the Lagrangian approach. In Refs.~\cite{ks1,ks2} we used the
homogeneous LC gauge as defined after (\ref{16}). Thus working
in LC gauges, $H_C$ need not be ``extended''. This would be
necessary in the conformal gauge ($q_3=\bar q_2 =0,\ q_2=\bar
q_3 =\exp(\ph)$) where $\dot{\bar q}_i \neq 0$ in general.

Disregarding the surface term, $H_C$ can be interpreted as a
generator of $G$-transformations with parameter $\bar q_i$. On
the constraint surface $H_C \sim 0$.

So far we discussed a LC formulation, because in terms of a
corresponding gauge the complete integrability of the equations
of motion is rather easy. The case of general axial-type gauges
--- without specifying in advance the space-time property of the
evolution parameter --- has been worked out as well \cite{strobl}.

\subsection{Evolution of Momenta}

The classical equations of motion for the momenta $\dot p_i = -
{\6 \CH_C \over \6 q_i}$ from (\ref{H})
\begin{eqnarray}
\label{p1}
\dot p_1 &=& - \bar q_3p_3 + \bar q_2p_2 \\
\label{p2}
\dot p_2 &=& - \bar q_1p_2 + \bar q_3 E \\
\label{p3}
\dot p_3 &=& \bar q_1p_3 - \bar q_2 E
\end{eqnarray}
for given gauge functions $\bar q_i(x^-)$ are closed for the momenta
alone and hold together with $G_i = 0$ (\ref{sc}). Moreover they are
``ultralocal'' in the sense that no space derivatives $\6_-$ appear.
The quantity $P$ (\ref{P}) is immediately recognized as a constant of motion
($\dot P = 0$). In addition, from (\ref{p2}) and (\ref{p3})
\begin{equation}
\label{211}
\6_+\left(p_2p_3\right) + E\dot p_1 =0
\end{equation}
follows which implies $\dot Q = 0$ with $Q$ already introduced
in (\ref{Q}). From the vanishing of $G_2$ and $G_3$ in
(\ref{sc}) an analogous relation
\begin{equation}
\label{212}
\6_-\left(p_2p_3\right) + E p_1^\pr =0
\end{equation}
can be obtained, proving that also $Q^\pr = 0$. Thus the result
(\ref{19}) of \cite{ks1} has been recovered that $Q$ is
conserved in space and time.

Eliminating $p_2$ and $p_3$ in (\ref{p1}) by $E$ and $Q$
according to (\ref{Q}) and (\ref{E}), and using $P$ from
(\ref{P})
\[
{\dot p_1}^2 = F(p_1)
\]
follows with
\begin{equation}
\label{213}
F = \left(P-\bar q_1p_1\right)^2 - 8\b \bar q_2\bar q_3
\left[ {{p_1}^2\over 2} - \l + 2\b^2 \left( {p_1\over \b} + 2
- \exp\left({p_1\over 2\b}\right) Q \right)
\right]\; ,
\end{equation}
which is readily integrated as long as $F>0$
\begin{equation}
\label{214}
x^+ = x^+(p_1,x^-) = \pm \int^{p_1} d\hat p_1 F^{-{1\over 2}}(\hat p_1)
+ \r(x^-)\; .
\end{equation}
The correct sign has to be fixed with help of Eq.~(\ref{p1}).
Having inverted (\ref{214}) for $p_1$, the momenta $p_2$ and
$p_3$ are given by the algebraic Eqs.~(\ref{Q}) and (\ref{P}).

In the homogeneous LC gauge $\bar q_1 = \bar q_2 = \bar q_3 - 1 = 0$
of \cite{ks1} the integrations are trivial and even independent of
$Q$
\begin{equation}
\label{216}
\begin{array}{rcl}
p_3 &=& P_0 \\
p_1 &=& -P_0\left( x^+ - \r(x^-) \right)\; ,
\end{array}
\end{equation}
where $P = P_0 = p_3$ in the homogeneous LC gauge. $p_2$ is
determined algebraically by Eqs.~(\ref{Q}) and (\ref{E}).
Comparing with Ref.~\cite{ks1}, the arbitrary functions $P_0$
and $\r$ are identified with $f$ and $h$ of that work:
\begin{equation}
\label{216a}
\begin{array}{rcl}
P_0 &=& 2\b f \\
\r &=& - h/f
\end{array}
\end{equation}

The case $P=0$ must be treated separately. In the homogeneous LC
gauge it implies $p_3 =  0$ (i.e.~vanishing torsion). As
$q_2=e\neq 0$, $p_1$ is fixed by the third Eq.~(\ref{sc}) to be
$p_1 = \pm\sqrt{2\l}$.  It exists for $\l\geq 0$.  Furthermore
$G_1=0$ requires $p_2=0$. This branch was called de Sitter
solution in \cite{ks1}.

Another interesting limit for the solutions of (\ref{p1}) --
(\ref{p3}) is obtained at $\b \to \infty$ for $p_2$ and $p_3\neq
0$.  In that case the quantity $E$ (\ref{E}) becomes independent
of $p_2p_3$ and, repeating the steps (\ref{211}) and (\ref{212}),
the corresponding conserved quantity $Q_{\infty}$ reads
\begin{equation}
\label{218}
Q_{\infty} = {{p_1}^3\over 6} - \l p_1 + p_2p_3\; .
\end{equation}
It may be obtained as well by expanding (\ref{Q}) in powers of $1/\b$
as
\[
Q = 2 - {\l \over 2\b^2} - {1 \over 4\b^3}Q_{\infty} +
\CO\left({1\over \b^4}\right)\; .
\]
The remaining steps are as above. The curvature varies in this case,
but the torsion scalar vanishes as $T^2\propto p_2p_3\b^{-2}$.
We shall call this branch of solutions the ``Einstein''
branch, because it contains nontrivial curvature with vanishing torsion,
reminiscent of Einstein gravity.
It should be noted that this limit does not commute with the procedure
leading to the equations of motion. In this sense it is a singular one.

\subsection{Evolution of Coordinates}

The Hamiltonian equations for $q_i$, $\dot q_i =
{\6 \CH_C \over \6 p_i}$ read:
\begin{eqnarray}
\label{219}
\dot q_1 &=& \bar q_1^\pr - e p_1 \\
\label{220}
\dot q_2 &=& \bar q_2^\pr + \bar q_1q_2 - \bar q_2q_1 + e\frac{1}{2\b}p_3 \\
\label{221}
\dot q_3 &=& \bar q_3^\pr - \bar q_1q_3 + \bar q_3q_1 + e\frac{1}{2\b}p_2
\end{eqnarray}
The dependence on space-derivatives is restricted to the
gauge-functions $\bar q_i$. Thus, for the purpose of dynamics,
Eqs. (\ref{219}) -- (\ref{221}) are as ultralocal as (\ref{p1})
-- (\ref{p3}). The integration at general $\bar q_i$, with $p_i$
as input from (\ref{214}), (\ref{P}) and (\ref{Q}) is simple. We
note first that the constraints $G_1=0$ and $G_3=0$ allow to
express $q_3$ and $q_1$ in terms of $q_2$
\begin{equation}
\label{223}
\begin{array}{rcl}
q_1 &=& (q_2E + p_3^\pr)/p_3 \\
q_3 &=& (q_2p_2 - p_1^\pr)/p_3\; .
\end{array}
\end{equation}
We introduce (\ref{223}) into (\ref{220}) and obtain a linear
inhomogeneous differential equation for $q_2$. E.g.~for
$\bar q_2=0$ its solution is given by
\begin{equation}
\label{224}
q_2(x^+,x^-)= \exp\left[ x^+\bar q_1 + \frac{\bar q_3 p_3(0,x^-)}{2\b}
\left(\frac{e^{\bar q_1x^+} - 1}{\bar q_1} \right) + \s(x^-) \right]\; .
\end{equation}
In the homogeneous LC gauge (\ref{224}) is further simplified
and becomes
\begin{equation}
\label{222}
e = q_2(x^+,x^-) = \exp\left( {P_0\over 2\b} x^+ + \s(x^-)\right)\; .
\end{equation}
In that case yet another method may be used, because then
\begin{equation}
\label{224a}
V_0 =q_2\exp\left( {p_1\over 2\b} \right)
\end{equation}
is readily identified as a constant of motion. It essentially
coincides with the arbitrary function $F\cdot f$ of
Ref.~\cite{ks1} (cf.~the explicit solution (\ref{231}) below).
Expressing $q_2$ by $p_1$ according to (\ref{224a}),
Eqs.~(\ref{223}) provide the (algebraic) expressions for $q_1$
and $q_3$.

In the de Sitter case $P=0$ described after Eq.~(\ref{216a}) above,
the determinant e,
Eq.~(\ref{222}), is independent of $x^+$
\begin{equation}
\label{225}
e=\exp\left(\s(x^-)\right)
\end{equation}
whereas, with $p_1 = \pm \sqrt{2\l}$ and $p_2=p_3=0$ from
(\ref{219})
\begin{equation}
q_1 = - e p_1 x^+ + \m (x^-)\; .
\end{equation}
Finally (\ref{221}) is trivially integrated
\begin{equation}
q_3 = \int^{x^+} d\hat x^+ q_1(\hat x^+) + \n(x^-)\; .
\end{equation}

In the Einstein branch ($\b \to \infty$) Eq.~(\ref{220}) yields,
working in the homogeneous LC gauge, again (\ref{225}) as for
the de Sitter case. The determination of $q_1$ and $q_3$,
however, again proceeds most suitably using (\ref{223}) at
$\b\to \infty$.

\subsection{Independent Functions of The Solution}

Six first order differential equations in $x^+$ minus three
constraints yield a dependence on three functions of $x^-$ in
the solution. This counting turns out to be correct for the
homogeneous LC gauge, where, in fact, all those relations are
independent. E.g.~for the general branch of the solution in that
gauge they are $\r,P_0$ and $\s$ or $V_0$. We show that these
functions can be essentially eliminated by appropriate residual
diffeomorphisms and Lorentz transformations which do not change
the LC gauge $\o_+=\bar q_1(x^-), e_+^{\ -}=\bar q_2(x^-),
e_+^{\ +}=\bar q_3(x^-)$. A similar argument can be found in
\cite{kat2} for the conformal gauge.  In a neighbourhood of a
point $x$, a diffeomorphism $\tilde x^\m=\tilde x^\m (x)$,
together with a Lorentz transformation $\L^a_{\ b}(x)$, induces
transformations (going back to the notation $e_\m^{\ a},\o_\m$
instead of (\ref{q}))
\begin{eqnarray}
\nonumber
\tilde \o_\m(\tilde x) &=& \frac{\6x^\n}{\6\tilde x^\m}\left(
\o_\n -\6_\n\a\right) \\
\label{229}
\tilde e_\m^{\ a}(\tilde x) &=& \L^a_{\ b}(x(\tilde x))\frac{\6x^\n}
{\6\tilde x^\m} e_\n^{\ b}\; ,
\end{eqnarray}
where for $a,b \in \{+,-\}$ $\L^a_{\ b}$ is diagonal with $\L^+_{\ +}
=(\L^-_{\ -})^{-1}=
\exp(\a(x))$. The residual transformations for $\tilde e_+^{\ +}=e_+^{\ +}
=1, \tilde e_+^{\ -}=e_+^{\ -}=\tilde \o_+=\o_+=0$, with
(\ref{229}) yield the restrictions
\begin{equation}
\label{230}
\begin{array}{rcl}
\a &=& \a(x^-) \\
x^- &=& \x^-(\tilde x^-) \\
x^+ &=& \exp(-\a)\tilde x^+ +\x^+(\tilde x^-)\; .
\end{array}
\end{equation}
In the homogeneous LC gauge the solution (\ref{223}), (\ref{222})
may be written in the notation of \cite{ks1} as (cf.~(\ref{lcinv}),
$\L = {\l\over 2\b^2}$ and $C_0 = 8\b Q$)
\begin{eqnarray}
e_-^{\ -} &=& Ffe^{\hat R} \nonumber \\
\label{231}
\o_- &=& 2\b Fe^{\hat R}\left( \hat R -1 \right) + \b QF
       +\frac{f^\pr}{f} \\
e_-^{\ +} &=& \frac{\b F}{f}e^{\hat R}\left( (\hat R -1)^2 +1 -\L\right)-
        \frac{\b QF}{f} + \frac{{\hat R}^\pr}{f} \nonumber
\end{eqnarray}
with
\begin{equation}
\label{232}
\hat R = fx^+ + h\; .
\end{equation}
The arbitrary functions of $x^-$ in (\ref{231}) and (\ref{232}) are
$F>0,f>0$ and $h$. Let us define a ``normal'' solution with
$F=f=1,h=0$ ($\hat R_0 = x^+$)
\begin{eqnarray}
e_-^{\ -} &=& e^{\hat R_0} \nonumber \\
\label{233}
\o_- &=& 2\b e^{\hat R_0}\left( \hat R_0 -1 \right) + \b Q \\
e_-^{\ +} &=& \b e^{\hat R_0}\left( (\hat R_0 -1)^2 +1 -\L\right) -
        \b Q \; . \nonumber
\end{eqnarray}
It is straightforward to show that the transformation of the dynamical
variables
\begin{eqnarray}
\nonumber
\tilde e_-^{\ -} &=& \exp(-\a)\frac{\6 x^-}{\6 \tilde x^-} e_-^{\ -} \\
\label{234}
\tilde e_-^{\ +} &=& \exp(\a)\left( \frac{\6 x^+}{\6 \tilde x^-}
+\frac{\6 x^-}{\6 \tilde x^-}e_-^{\ +} \right) \\
\nonumber
\tilde \o_- &=& \frac{\6 x^-}{\6 \tilde x^-}\left( \o_- - \6_-\a \right)
\end{eqnarray}
with (\ref{230}) and (\ref{233}) reproduces the general
expression (\ref{231}) with $\tilde F, \tilde f$ and $\tilde h$
given in terms of $\a$ and $\x^\pm$ and their first derivatives.
{}From the solution of the equation of motion and of the
constraints leading to (\ref{231}) \cite{ks1}, $\tilde F,\tilde
f$ and $\tilde h$ must allow at least one differentiation. The
class of functions produced by $\a$ and $\x^\pm$ is certainly
larger. Therefore (once differentiable) functions $\tilde
F,\tilde f$ and $\tilde h$ can be always produced.

For the de Sitter solution, again in the notation of \cite{ks1},
\begin{eqnarray}
e_-^{\ -} &=& q(x^-) \nonumber \\
\label{235}
\o_- &=& \frac{R_0}{4} q x^+ + l \\
e_-^{\ +} &=& \frac{R_0}{8}q{x^+}^2 + lx^+ +s
\nonumber
\end{eqnarray}
with $R_0 = \pm 8\b \sqrt{\L}$. The ``normal'' solution may be
chosen as
\begin{eqnarray}
e_-^{\ -} &=& 1 \nonumber \\
\label{236}
\o_- &=& \frac{R_0}{4}x^+ \\
e_-^{\ +} &=& \frac{R_0}{8}{x^+}^2\; .
\nonumber
\end{eqnarray}
Again applying (\ref{234}) with (\ref{236}) produces (\ref{235}) in
terms of variables $\tilde x^+,\tilde x^-$ with (${dX(\tilde x^-)\over
d\tilde x^-} = X^\pr$, $\tilde \a = \a(x(\tilde x))$)
\begin{eqnarray}
\tilde q &=& \exp(-\tilde \a){\x^-}^\pr \nonumber \\
\label{237}
\tilde l &=& -\tilde \a^\pr + {R_0\over 4} \x^+ {\x^-}^\pr \\
\tilde s &=& \exp(\tilde \a)\left( {\x^+}^\pr + {R_0\over 8}
{\x^+}^2 {\x^-}^\pr \right)\; .
\nonumber
\end{eqnarray}
Our argument, of course, implies that the inverse transformation can be
performed as well in local patches, i.e.~that $x^-$ can be eliminated
from the general solution in regions bounded by coordinate
singularities.

The present argument works for the homogeneous LC gauge and with
the explicit solutions.  It is equally true in more general
gauges (cf.~the proof for the conformal gauge \cite{kat1}).  We
checked it in the case where $\bar q_1 \neq 0, \bar q_3 \neq 0$
but $\bar q_2 = 0$. (\ref{230}) is replaced by
\begin{eqnarray}
\nonumber
\a(x^+,x^-) &=& - \ln\left[  1 + \exp(-\bar q_1 x^+)
\left( \exp(-\a) - 1 \right) \right] \\
\label{230a}
x^+ &=& \frac{1}{\bar q_1}\ln\left[ \exp\left(\bar q_1\tilde x^+ +
\bar q_1B(x^-) - \a \right) + 1 - \exp(-\a) \right] \\
\nonumber
x^- &=& \x^-(\tilde x^-)\; ,
\end{eqnarray}
where $\a=\a(0,x^-)$. In the limit $\bar q_1 \to 0$ we obtain
(\ref{230}) by identifying $B=\x^+\exp(\a)$.  It is then
straightforward but tedious to verify that there exists locally
a transformation of the form of Eq.~(\ref{229}) which maps two
solutions of the equations of motions to the same $Q$, differing
by the initial conditions, into each other.

The ultralocal structure of the canonical equations of motion
agrees with the fact that the fields at each point $x^-$ in
space evolve independently (in the homogeneous LC gauge). Thus
for all times by having fixed one transformation of space the
latter may be expressed in a trivial manner. Having eliminated
$x^-$ altogether from the solution of (\ref{p1}) -- (\ref{p3})
and (\ref{219}) -- (\ref{221}), with vanishing constraints
(\ref{sc}) and costant $\bar q_i$, we may conclude that, in
order to understand the dynamics of this system, it is enough to
consider a phase space $p_i(x^+), q_i(x^+)$, i.e.~to regard
(\ref{p1}) -- (\ref{p3}) and (\ref{219}) -- (\ref{221}) at
$\bar q_i=const$ with (\ref{sc}) at $p_i^\pr = 0$ as the
dynamics of a three dimensional point particle with time $x^+$.
It may be related to a Hamiltonian $H^{eff} =
-\bar q_iG_i|_{p_i^\pr=0}$, with Poisson brackets of a point
particle.  Strictly speaking, also here $\bar q_i$ have to be
interpreted as Lagrange-multipliers to be fixed by a gauge
condition.

This simplification of the solutions allows too a concise
discussion of the different possibilities of nontrivial
evolution in $x^+$. The expressions for the curvature scalar and
the squared torsion with (\ref{233}) read ($R=8\b\hat R,
T^2=8\b\hat T^2$):
\begin{eqnarray*}
\hat R &=& x^+ \\
\hat T^2 &=& Q \exp \left( -x^+ \right) - \left( x^+ - 1 \right)^2
+ \L - 1
\end{eqnarray*}
Starting, e.g. from $x^+=0$ where
\begin{equation}
\label{238}
\begin{array}{rcl}
\hat R|_{x^+=0} &=& 0 \\
\hat T^2|_{x^+=0} &=& Q + \L -2\; ,
\end{array}
\end{equation}
the curvature always changes linearly with a singularity at
$x^+\to\infty$. Dynamically different branches of the general
solution are simply classified by the value of $Q$ and of $\L -
1$ which determine the behaviour of $\hat T^2$.  We have at
$Q=0$ the possibilities for the evolution of $\hat T^2$
\begin{equation}
\label{239}
\begin{array}{rcl}
\L<1 &\then& \hat T^2<0 \\
\L=1 &\then& \mbox{one zero of } \hat T^2 \\
\L>1 &\then& \mbox{two zeros of } \hat T^2,
\end{array}
\end{equation}
whereas for $Q<0$
\begin{equation}
\label{240}
\begin{array}{rcl}
\L\leq 1 &\then& \hat T^2<0 \\
\L>1 &\then& \mbox{two zeros of } \hat T^2.
\end{array}
\end{equation}
If $Q>0$ the number of zeros of $\hat T^2$ depends on the number of
intersections of an exponential with a parabola ($\leq 3$
zeros).

Thus, having $Q$ fixed by one initial value condition, e.g.~at
the origin of the coordinate system, for a given set of
parameters in (\ref{inv}) the dynamics of the system is fully
determined.

\section{Symmetry Generated by Secondary Constraints}

\subsection{Algebra of Constraints}

The Poisson brackets (\ref{pb}) of the secondary constraints $G_i$
(\ref{sc}) yield
\begin{equation}
\label{31}
\begin{array}{rcl}
[G_1,G_2] &=& - G_2 \d \\
{[}G_1,G_3] &=& G_3\d \\
{[}G_2,G_3] &=& - \left[ p_1G_1 -
                {1\over 2\b} \left( p_3G_2+p_2G_3 \right) \right]\d\; ,
\end{array}
\end{equation}
where $\d(x^- - y^-)$ has been abbreviated by $\d$. (\ref{31}) implies the
absence of ternary constraints, used already above. The generator of
local $G$-transformations
\begin{equation}
\label{32}
G_\x = \int dx^- \z_i(x^-)G_i(x^+,x^-)
\end{equation}
induces for any $\z_i(x^-)$ variations of a variable $F(p,q)$
\begin{equation}
\label{33}
\d F= [F,G_\z]\; .
\end{equation}
If we disregard surface effects, assuming appropriate boundary
conditions (or periodicity) so that the total divergence $P^\pr$
in (\ref{H}) may be dropped, the Hamiltonian coincides with
(\ref{32}) for the identification $\z_i \to \bar q_i$ (the gauge
functions).

(\ref{33}) may be evaluated for any $F$ in terms of $[G_i,p_j]$
\begin{equation}
\label{34}
\begin{array}{r@{}c@{}c@{}c@{}lcl}
[ & G_{\underline{i}} &,& p_{\underline{i}} & ] &=& 0 \\
{[} & G_1 &,& p_2 & ] &=& [p_1,G_2] = - p_2\d \\
{[} & G_1 &,& p_3 & ] &=& [p_1,G_3] =   p_3\d \\
{[} & G_2 &,& p_3 & ] &=& [p_2,G_3] = - E\d
\end{array}
\end{equation}
and of $[q_i,G_j]$
\begin{equation}
\label{35}
\begin{array}{rcl}
[q_1,G_1] &=& - \d^\pr \\
{[}q_1,G_2] &=& - q_3p_1\d \\
{[}q_1,G_3] &=& q_2p_1\d \\
{[}q_2,G_1] &=& -q_2\d \\
{[}q_2,G_2] &=& -\d^\pr +\left({2\over \b}q_3p_3 + q_1\right)\d \\
{[}q_2,G_3] &=& -{2\over \b}q_2p_3\d \\
{[}q_3,G_1] &=& q_3\d \\
{[}q_3,G_2] &=& {2\over \b}q_3p_2\d \\
{[}q_3,G_3] &=& -\d^\pr -\left({2\over \b}q_2p_2+q_1\right)\d
\end{array}
\end{equation}
where $\d^\pr = \6_-\d(x^- - y^-)$. The $G$-algebra (\ref{31}) is linear
in $G$, with $p$-dependent structure functions on the r.h.s. It is in weak
involution, but not closed and resembles the algebra of secondary constraints
in general relativity using the Ashtekar formulation with polynomial
$G_i$ \cite{ash}. In contrast to that case, however, a
 minimal closed algebra involving (\ref{31}) is obtained
taking the ``G-p-algebra'' (\ref{31}) and (\ref{34}). The corresponding
minimal set of elements for which this algebra is closed is
\begin{equation}
\label{36}
A=A_0(p)+A_i(p)G_i\; ,
\end{equation}
where $A_0$ and $A_i$ are polynomials (or analytic functions or formal
power series) in $p_i$. A minimal requirement for the $A(p)$'s is that
they should be elements of a commutative ring containing a unit
element. The algebra of elements (\ref{36}), endowed
with the brackets (\ref{31}) and (\ref{34}) of the G-p-algebra is
ultralocal and shall be denoted as $\CA$.

In terms of the new basis
\begin{equation}
\label{37}
\begin{array}{rcl}
X_0 &=& G_1 \\
X_{\mp} &=& {1\over |2\b|^{1/2}} G_{2\atop 3} \\
1 + \n Z_0 &=& - {1\over 2\b}p_1 \\
\n Z_{\mp} &=& - {1\over 2\b|2\b|^{1/2}}p_{2\atop 3}
\end{array}
\end{equation}
the defining relations of the G-p-algebra are particularly
transparent if the quadratic form of $iso(2,1)$
\begin{equation}
\label{38}
\la Y,X \ra = Y_0X_0 - \mbox{sgn}\b \left( Y_+X_- + Y_-X_+ \right)
\end{equation}
is introduced ($a,b = 0,+,-$):
\begin{equation}
\label{39}
\begin{array}{r@{}c@{}c@{}c@{}lcl}
[&X_0&,&X_{\pm}&] &=& \pm X_{\pm} \d \\
{[}&X_+&,&X_-&] &=& - \mbox{sgn}\b\left( X_0 + \n \la Z,X \ra \right) \d \\
{[}&X_{\underline{a}}&,&Z_{\underline{a}}&] &=& 0 \\
{[}&X_0&,&Z_{\pm}&] &=& [Z_0,X_{\pm}] = \pm Z_{\pm}\d \\
{[}&X_-&,&Z_+&] &=& [Z_-,X_+] = \mbox{sgn}\b\left( Z_0 +{\n\over 2}
\la Z,Z \ra - {\L - 1\over 2\n} \right) \d \\
{[}&Z_a&,&Z_b&] &=& 0
\end{array}
\end{equation}
In Eqs.~(\ref{39}) we recognize an $iso(2,1)$ Lie-Poisson
algebra, deformed by $\n$-dependent terms. The deformation is
quadratic and is identified with the Casimir operators of an
undeformed $iso(2,1)$ plus a central term depending on the
cosmological constant.

The de Sitter solution corresponds (for $\l\geq 0$) to the orbit
$p_1=\pm\sqrt{2\l}$, $p_2=p_3=0$. On this orbit the algebra
(\ref{31}) is itself closed: for $\l>0$ we recognize the algebra
$sl(2,\re)$ and for $\l = 0$ the two dimensional Poincar\'e algebra
$iso(1,1)$ is recovered. This is as expected,
because the de Sitter solution is related to the Liouville
equation \cite{kat1} and $sl(2,\re)$ plays a central r\^ole in
Liouville theory \cite{dhooker}.

The Einstein solution ($\b\to \infty$) possesses an algebra
which can be read off from (\ref{31}) and (\ref{34}), simply
dropping the quadratic terms with $p_2p_3, G_2p_3$ and $p_2G_3$
on the respective r.h.s. The only remaining quadratic terms are
$p_1G_1$ and $p_1^2$, i.e.~involve mutually commuting elements.
Nonlinear generalizations of Lie algebras involving presicely
such quadratic expressions of the elements of the Cartan
subalgebra have been considered already in the literature
\cite{nss}. Here an example for this idea emerges in a natural
manner as the dynamical symmetry of a specific model.

\subsection{Centre of $\CA$}

We call the functions (\ref{36}) which commute with all
generators $F_\a =(G_i,p_j)$ ``elements of the centre'' of
$\CA$. In order to determine their number, it is convenient to
consider the completely gauge-fixed situation of Sec.~2.4 where
the $x^-$-dependence has been eliminated. It is then sufficient
to consider the algebra (\ref{31}) and (\ref{34}) also with
respect to this six-dimensional phase space only. Obviously the
elements of the centre are solutions of
\begin{equation}
\label{310}
M_{\a\b}{\6 A\over \6 F_\b} = 0\; ,
\end{equation}
where
\begin{equation}
\label{311}
M_{\a\b} = [F_\a,F_\b]\; .
\end{equation}
The antisymmetric matrix $M_{\a\b}$ consists of the blocks
$[G,G],[G,p], [p,G]$ and $[p,p]$ with (\ref{31}) and (\ref{34}),
without delta-functions on the r.h.s. It is easy to see that
$M_{\a\b}$ has rank four, i.e.  that there are at most two
independent elements $A$ solving (\ref{310}) in a nontrivial
manner.\footnote{The theory of elements $A$ of a Poisson algebra
$\CA$ has been developped a long time ago by Engels and Lie
under the title ``function groups'', the elements of the centre
were called ``singular elements'' in this context. A short but
comprehensive introduction into this work and further references
are provided by \cite{eis}.} The two solutions of (\ref{310})
are $Q$ and $K$ with $Q$ given by (\ref{Q}), and
\begin{equation}
\label{312}
K = {1\over 4\b^2} \exp\left( -{p_1\over 2\b} \right) \left[ E G_1
+ p_3G_2 + p_2G_3 \right]
\end{equation}
as can be verified easily. Whereas $Q$ has appeared already in
the solution of the classical problem, (\ref{312}) is new, but
it vanishes identically on the constraint surface $G_i=0$. From
the explicit expression (\ref{H}), it is evident that both $Q$
and $K$ commute with $H_C$. In terms of the rescaled basis
(\ref{37}), $Q$ and $K$ read
\begin{eqnarray}
\label{313}
Q &=& \n^2 \exp(1 + \n Z_0)
      \left( \la Z,Z \ra - {\L - 1\over \n^2} \right) \\
\label{314}
K &=& \n \exp(1 + \n Z_0) \la Z,X \ra + {1\over 2} Q X_0.
\end{eqnarray}
The general proof for $x^-$-dependent $M_{\a\b}$ is straightforward
too. Every $F(X_a,Z_a) \in \CU (\CA)$ which commutes with $X_0$ may be
written as
\[
F=F(X_0,Z_0,Q,K,X_+X_-,X_+Z_- - X_-Z_+)\; .
\]
Evaluating $[X_\pm,F]=0$ and $[Z_a,F]=0$ one shows easily that $F=F(Q,K)$.

The algebra (\ref{39}) for $\n\to 0$ and $\L=1$ reduces to the
linear algebra $iso(2,1)$ and the nonpolynomial expressions
(\ref{313}) and (\ref{314}) become its Casimir operators. For
$\L\neq 1$ during this limit at one point a redefinition of
$p_1$ in Eq.~(\ref{37}) with a factor $\sqrt{\L}$ is necessary
in order to arrive at the same result.

In the Einstein limit $\b\to\infty$, the matrix $M_{\a\b}$
simplifies considerably, because terms with $p_2$ and $p_3$ are
dropped. Again its rank is four and the two elements are
$Q_\infty$ of (\ref{218}) and
\begin{equation}
\label{315}
K_\infty = E_{\infty} G_1 + p_3G_2 + p_2G_3\; ,
\end{equation}
where $E_{\infty} = {p_1^2\over 2} - \l$.

\section{Conclusions}

We have shown that the reason for the complete integrability of
$R^2$-gravity with torsion in $d=2$ is a new dynamical symmetry
of Hamiltonian phase space. This symmetry can be written as a
(nonlinear) closed Lie-Poisson algebra with the nonlinear
deformation proportional to Casimir operators of undeformed
$iso(2,1)$ and with a central extension involving the
cosmological constant. In the de Sitter limit the deformation
vanishes. For another special limit with vanishing torsion but
with nontrivial curvature the deformation reduces to quadratic
terms of the Cartan subalgebra of $iso(2,1)$.

Among the (two) elements of the centre of the new algebra we
recovered the locally conserved quantity, encountered previously
in the course of the explicit solutions, whereas the second
element vanishes identically on the constraint surface.  Fixing
the homogeneous LC gauge completely allows the reduction to a
certain mechanical system of a point particle in three
dimensions, depending on time alone.

The novel symmetry is a Lie-Poisson algebra and fulfils Jacobi's
identities. Replacing Poisson-brackets by commutators at fixed
$x^+$ in (\ref{pb})
\[
[\ ,\ ]_{Poisson} \to -i[\ ,\ ]_{commutator}
\]
and using Heisenberg equations of motion $\dot p_i = i[H_C,p_i]$
etc.~should lead to a quantized version of the present model.
The notorious problem of locality on a light-like plane in the
presence of massless excitations precludes such a step without
further analysis which is outside the scope of our present work
\cite{strobl,haider}. Disregarding this complication, we still
have to deal with operator ordering ambiguities in the
Hamiltonian. It turns out that the only potentially dangerous
terms are of the type $p_{\underline{i}}q_{\underline{i}}$, to
be simply defined as ${1\over2}\{p_{\underline{i}},
q_{\underline{i}}\}$ in order to obtain a hermitian $H_C$. With
correspondingly ordered operators in (\ref{sc}) the G-p-algebra
(Sec.~3) can be viewed alternatively as a nonlinear
commutator-algebra.

As a consequence of this Lie-Poisson algebra $\CA$, $R^2$-gravity
with torsion in $d=2$ possesses a dynamical symmetry, which
leads to a set of solutions which, in contrast to Einstein
gravity in d=2, are not just topological.  On the other hand,
the presence of such a symmetry improves the changes of
renormalizability and perhaps finiteness, if such a symmetry can
be maintained when further fields are added.

It should be emphasized again that this symmetry is visible at
the Hamiltonian level. As far as the action is concerned, it has
been shown that no further continuous symmetry beside the
original ones (diffeomorphisms plus local Lorentz
transformations) are present in this model \cite{strobl}.

Apart from possible applications in physics, the nonlinear
Lie-Poisson algebra, encountered in the context of this model
may also be suggestive for attempts deforming Lie algebras other
than $iso(2,1)$.

Among the open problems the prominent ones are quantization and
the treatment of finite boundaries which is closely related to
nontrivial topologies of the solution. This question is
especially important because in d=2 Green functions at large
distances are not damped naturally.

{\bf Acknowledgement:} The authors appreciate discussions with Th.~Strobl.

\end{document}